%% file: root.tex
\pdfoutput=1
\def\compileforpublish{1}
\def\isaccepted{1}

\documentclass[letterpaper, 10 pt, conference]{ieeeconf}  

\IEEEoverridecommandlockouts                              

\input{macros}

\usepackage{todonotes}
\usepackage{url}
\usepackage{graphicx}
\usepackage{subfig}

\title{\LARGE \bf
From Functional to Logical Scenarios: Detailing a Keyword-Based Scenario Description for Execution in a Simulation Environment
}%
\author{Till Menzel$^{1}$ and Gerrit Bagschik$^{1}$ and Leon Isensee$^{2}$ and Andre Schomburg$^{2}$ and Markus Maurer$^{1}$
\thanks{$^{1}$Authors are with the Institute of Control Engineering, 
		Techni\-sche Universit\"at Braunschweig, 38106 Braunschweig, Germany
        {\tt\small menzel, bagschik, maurer @ifr.ing.tu-bs.de}}%
\thanks{$^{2}$Authors wrote their master theses at the Institute of Control Engineering, 
		Techni\-sche Universit\"at Braunschweig, 38106 Braunschweig, Germany
		{\tt\small l.isensee, a.schomburg @tu-braunschweig.de}}%
}

\begin{document}
\maketitle%
\thispagestyle{empty}
\pagestyle{empty}

\begin{abstract}
	\input{abstract}

\end{abstract}

\vspace{-0.5cm}
\copyrightnotice

\input{introduction}

\input{relatedWork}

\input{specification}

\input{evaluation}

\input{conclusion}


\bibliographystyle{IEEEtran}
\bibliography{bib/bib}

\end{document}

%% file: macros.tex
\makeatletter
\newcounter{IEEE@bibentries}
\renewcommand\IEEEtriggeratref[1]{%
	\renewbibmacro{finentry}{%
		\stepcounter{IEEE@bibentries}%
		\ifthenelse{\equal{\value{IEEE@bibentries}}{#1}}
		{\finentry\@IEEEtriggercmd}
		{\finentry}%
	}%
}
\makeatother

\ifx \isaccepted \undefined
\newcommand\copyrighttext{%
	\footnotesize \centering This work has been submitted to the IEEE for possible publication.\\ Copyright may be transferred without notice, after which this version may no longer be accessible.}
\else
\newcommand\copyrighttext{%
	\footnotesize \parbox[t]{.11\textwidth}{\copyright{} \the\year~IEEE.} \parbox[t]{.89\textwidth}{Personal use of this material is permitted. Permission from IEEE must be obtained for all other uses, in any current or future media, including reprinting/republishing this material for advertising or promotional purposes, creating new collective works, for resale or redistribution to servers or lists, or reuse of any copyrighted component of this work in other works.}}
\fi
\newcommand\copyrightnotice{%
	\ifx \compileforpublish \undefined
	\else
	\begin{tikzpicture}[remember picture,overlay]
	\node[anchor=south,yshift=10.5pt] at (current page.south) {\parbox{\dimexpr\textwidth-\fboxsep-\fboxrule\relax}{\copyrighttext}};
	\end{tikzpicture}%
	\fi
}



%% file: abstract.tex
Scenario-based development and test processes are a promising approach for verifying and validating automated driving functions. 
For this purpose, scenarios have to be generated during the development process in a traceable manner. 
In early development stages, the operating scenarios of the item to be developed are usually described in an abstract, linguistic way. 
Within the scope of a simulation-assisted test process, these linguistically described scenarios have to be transformed into a state space representation and converted into data formats which can be used with the respective simulation environment. 
Currently, this step of detailing scenarios takes a considerable manual effort. 
Furthermore, a standardized interpretation of the linguistically described scenarios and a consistent transformation into the data formats are not guaranteed due to multiple authors as well as many constraints between the scenario parameters.
In this paper, the authors present an approach to automatically detail a keyword-based scenario description for execution in a simulation environment and provide a basis for test case generation. 
As a first step, the keyword-based description is transformed into a parameter space representation. 
At the same time, constraints regarding the selection and combination of parameter values are documented for the following process steps (e. g. evolutionary or stochastic test methods). 
As a second step, the parameter space representation is converted into data formats required by the simulation environment. 
As an example, the authors use scenarios on German freeways and convert them into the data formats OpenDRIVE (description of the road) and OpenSCENARIO (description of traffic participants and environmental conditions) for execution in the simulation environment Virtual Test Drive.

%% file: introduction.tex
\section{Introduction}

Before introducing automated driving functions, a specified safety level has to be guaranteed.
For driver assistance systems, this is tested by a distance-based approach.
For systems of higher level automation, Wachenfeld and Winner~\cite{WachenfeldRelease2016} state 6.62 billion test kilometers to hypothetically prove (with a 50\,\% chance) that these systems are twice as good as a human driver.
Consequently, Wachenfeld and Winner conclude that for the release of systems of higher level automation a distance-based approach is economically not acceptable.
Therefore, Schuldt \cite{schuldt_beitrag_2017} suggests a scenario-based approach as a promising alternative.

\begin{figure*}[t]
	\centering
	\includegraphics[width=0.90\textwidth]{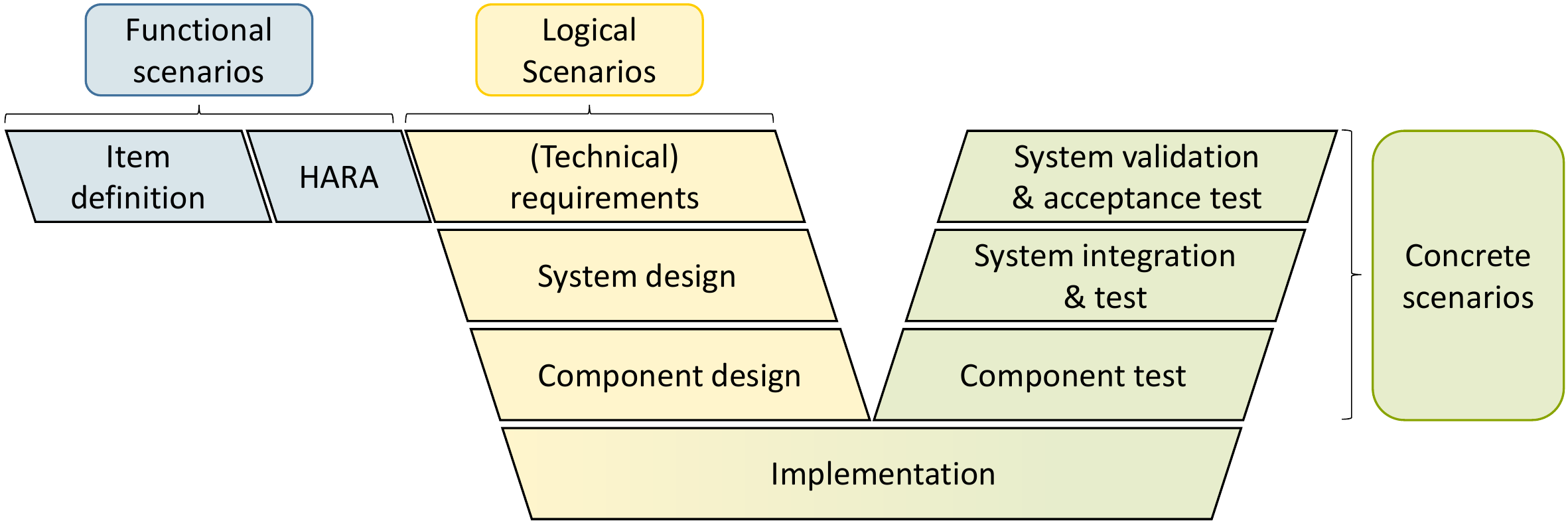}
	\caption{Scenarios during a V-model-based development process \cite{SzenarienProzess2017}; HARA: hazard analysis and risk assessment}
	\label{fig:V-model}
\end{figure*}

In scenario-based approaches, the scenario generation is a sensitive step for the safety argumentation and, thus, critical for the release of the system.
Hence, the scenarios have to be derived and documented systematically.
In addition, they have to be traceable along the development process.
For this purpose, Menzel et al. \cite{menzel_process_2018} propose three levels of abstraction for scenario representation: functional scenarios, logical scenarios and concrete scenarios.
Functional scenarios are described in a linguistic way so that experts can talk about scenarios in the beginning of the development process.
Logical scenarios specify parameters for the scenarios and define parameter ranges.
Concrete scenarios specify a concrete value for each parameter and, thus, are the basis for reproducible test cases.
Hence, these levels of abstraction support the activities during the concept phase, the system development as well as the testing activities of a V-model-based development process and build the basis for a structured generation and use of scenarios during the development process, as shown in Fig.~\ref{fig:V-model}.

For an automated generation of functional scenarios, Bagschik et al. \cite{bagschik_Szenariengenerierung_2018} present a knowledge-based approach.
The generated scenarios are described using a defined vocabulary and grammar.
Each scenario consists of a start scene and an end scene.
For the transformation from functional to logical scenarios, the keyword-based representation has to be detailed in a parameter space representation and converted into the formats for the simulation environment.
Within the scope of a sequence control for simulation, this step also includes the specification of the interactions of the traffic participants between the start and the end scene.
To consistently model the road and the interactions of the traffic participants according to the data formats, relations and dependencies between the parameters and parameter values have to be taken into account.

Currently, the steps of scenario detailing and scenario conversion require a huge manual effort. 
Furthermore, especially in case of multiple authors, a standardized interpretation of linguistic scenarios and a consistent representation compliant with the data formats for the simulation environment is not ensured.
In this paper, an approach for an automated transformation from a keyword-based scenario description to a parameter space representation and a conversion into the formats for a simulation environment is presented using the example of scenarios on German freeways.
Within the parameter space representation, relations and dependencies of the elements and the parameters of the scenario are modeled.
This is required to ensure that the interactions of the traffic participants are resolved according to the specification of the respective functional scenario and to ensure a consistent implementation of the scenario compliant with the data formats.  
For the representation of the road network, the data format OpenDRIVE is used.
For the representation of the traffic participants and the environment, the data format OpenSCENARIO is used.

This paper is based on an earlier German version \cite{menzel_detaillierung_2018}; it is structured as follows: Section \ref{related_work} provides the motivation for a knowledge-based scenario generation and scenario detailing based on selected related work.
Section \ref{sec:specification} presents an approach for detailing keyword-based scenarios for an execution in a simulation environment and as a basis for generating test cases.
In Section \ref{sec:evaluation} the implemented tool for detailing scenarios is evaluated.
Finally, Section \ref{sec:conclusion} shows some limitations of the presented approach and gives a short conclusion.

%% file: relatedWork.tex
\section{Related Work}
\label{related_work}

Schuldt et al. \cite{schuldt_2013} suggest a scenario-based test approach as well as a 4-layer-model for structuring scenarios.
Definitions of the terms scene, situation and scenario are provided by Ulbrich et al. \cite{ulbrich_definition_2015}.
Menzel et al. \cite{menzel_process_2018} use theses definitions as a basis and motivate different notations for representing scenarios during the development process using the example of the standard ISO 26262 \cite{ISO_26262_2016}.
Consequently, Menzel et al. suggest three levels of abstraction for scenarios: functional scenarios, logical scenarios and concrete scenarios.
In this contribution, the terminology according to the definitions of Ulbrich et al. \cite{ulbrich_definition_2015} and Menzel et al. \cite{menzel_process_2018} is used.

\begin{figure*}[htbp]
	\centering
	\includegraphics[width=0.90\textwidth]{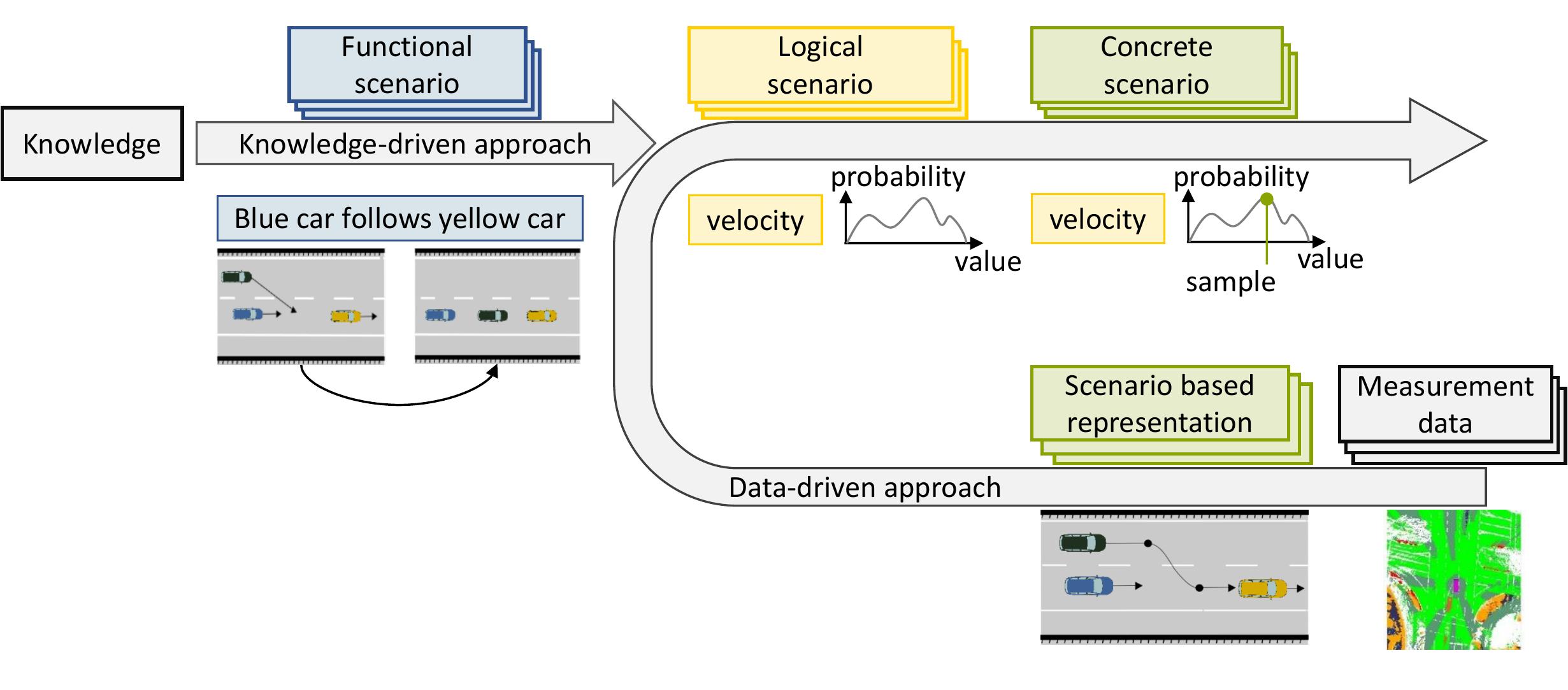}
	\caption{Comparison of a knowledge-driven and a data-driven approach for scenario generation.}
	\label{fig:approaches}
\end{figure*}

Currently, there are two approaches to the process of generating scenarios.
Using the terminology by Menzel~et~al.~\cite{menzel_process_2018}, both approaches can be compared, as shown in Fig.~\ref{fig:approaches}.
On the one hand, scenarios can be generated with the help of a data-driven approach, on the other hand, scenarios can be generated with the help of a knowledge-driven approach.

The main idea of a data-driven approach is to collect measurement data and identify as well as classify occurring scenarios.
There are multiple contributions which follow a data-driven approach, for example \cite{gelder_assessment_2017}, \cite{puetz_database_2017} and \cite{waymo_2017}.
De~Gelder~et~al.~\cite{gelder_assessment_2017} present an approach to generate realistic test cases for virtual testing like hardware-in-the-loop testing.
For this purpose, de Gelder et al. collect real-life scenarios, parameterize these scenarios and store them in a database.
For the real-life scenarios, it is assumed that they can be completely constructed from sensory and external information like digital maps and weather data.
The parametrization is shown exemplarily.
Which parameters are required for representation of the respective scenario is not described in detail.
After storing the parameterized scenarios in a database, a distribution is fit to model the scenario parameters.
This representation describes the parameter space to be tested, including the probability density function for each parameter.
Following the terminology by Menzel et al. \cite{menzel_process_2018}, this notation is equivalent to a logical scenario.
According to the distributions modeled in the logical scenarios, de Gelder et al. sample new (concrete) scenarios and execute them in a simulation environment.
This sampling forms the basis of a Monte Carlo Simulation.
Nevertheless, the approach by de Gelder et al. and similar approaches are classified as data-driven processes because all generated scenarios are based on real-life scenarios and no semantic variation of scenarios is performed.

P\"utz et al. \cite{puetz_database_2017} present a similar approach that describes a concept for a tool chain to collect data (for example measurement data and traffic simulation data), calculate metrics to characterize scenarios and cluster them as logical scenarios.
These logical scenarios are provided as a basis for generating test cases.

Waymo LLC \cite{waymo_2017} presents a test approach in which parts of recorded test drives can be reproduced in simulation.
For example, parameters describing traffic participants can be varied.
For this approach, measurement data of the test drives as well as a highly detailed digital map of the road network is needed.
In a first step, this approach converts measurement data into a scenario-based representation which can be executed in a simulation environment.
In a second step, these scenarios are generalized to logical scenarios for a subsequent variation of parameters and, thus, for a generation of concrete scenarios respective test cases.

These three approaches from de Gelder et al., P\"utz et al. and Waymo LLC follow the idea of a data-driven test process within the scope of virtual testing.
A generation of synthetic roads or scenarios based on formalized knowledge is not described.
For those purely data-driven approaches, there are several challenges.
First of all, measurement data do not describe all aspects of a scenario.
In contrast to the assumption of a complete scenario construction from sensory and external data by de Gelder et al. \cite{gelder_assessment_2017}, there will be elements in the scenario that will not be perceived by the sensors of the measuring vehicle (for example due to range limitations).
Additionally, when summarizing scenario-based representations as logical scenarios, further knowledge has to be taken into account to specify parameter dependencies.
For example, in an overtaking scenario, different velocities of the vehicles can be chosen, but the velocity of the overtaking vehicle always has to be faster than the velocity of the overtaken vehicle.
Within the scope of test case generation, those constraints have to be taken into account when generating concrete scenarios based on logical scenarios.
Last but not least, data-driven approaches give no information of what has been encountered from the operational design domain so far.
Those approaches only cluster scenarios the test vehicle has encountered and vary the parameter values.
A reference to all possible clusters (semantic variation) according to the operational design domain is not given.
Hence, those data-based approaches provide no information regarding the diversity of scenarios.

To compensate the stated problems, data-driven approaches can be combined with a knowledge-driven approach.
The main idea is to generate scenarios from existing knowledge, like regulatory recommendations, and later augment these synthetic scenarios with information collected from measurement data.
As a first step, to generate functional scenarios, the available knowledge has to be structured, varied on a semantic level and described linguistically.
Therefore, Bagschik et al. \cite{bagschik_ontology_2018} present a know\-ledge-based approach to generate start scenes using an ontology.
To structure the know\-ledge represented in the ontology, Bagschik et al. use a 5-layer-model based on the 4-layer-model by Schuldt~et~al.~\cite{schuldt_2013}.
In a later publication, Bagschik~et~al.~\cite{bagschik_Szenariengenerierung_2018} extend the concept of the knowledge-based generation of start scenes and present an implementation for a knowledge-based generation of operating scenarios for German freeways.
Each operating scenario, which includes a start scene and an end scene, is linguistically described on a semantic level (functional scenario) and represented using the Web Ontology Language (OWL).
To use these scenarios as a basis for generating concrete scenarios within the scope of virtual testing, the functional scenarios generated by Bagschik et al. have to be transformed into a notation which is executable in a simulation environment.

For the simulative test of automated driving functions, a broad variety of simulation environments like ASM Traffic\footnote{\url{https://www.dspace.com/de/gmb/home/products/sw/automotive_simulation_models/produkte_asm/asm_traffic.cfm},  accessed on 30. January 2019} of dSPACE GmbH, CarMaker\footnote{\url{https://ipg-automotive.com/products-services/simulation-software/carmaker/}, accessed on 30. January 2019} of IPG Automotive GmbH or Virtual Test Drive\footnote{\url{https://vires.com/vtd-vires-virtual-test-drive/}, accessed on 30. January 2019} of VIRES Simulationstechnologie GmbH are available.
These simulation environments utilize different data formats to specify scenarios.

The tool presented in this paper transforms functional scenarios into the data formats for the simulation environment Virtual Test Drive.
For the description of the road network, Virtual Test Drive requires the data format OpenDRIVE\footnote{\url{http://www.opendrive.org/}, accessed on 30. January 2019}.
For the description of the traffic participants and the environment, Virtual Test Drive utilizes the event-based data format OpenSCENARIO\footnote{\url{http://www.openscenario.org/}, accessed on 30. January 2019}.

\begin{figure*}[h]
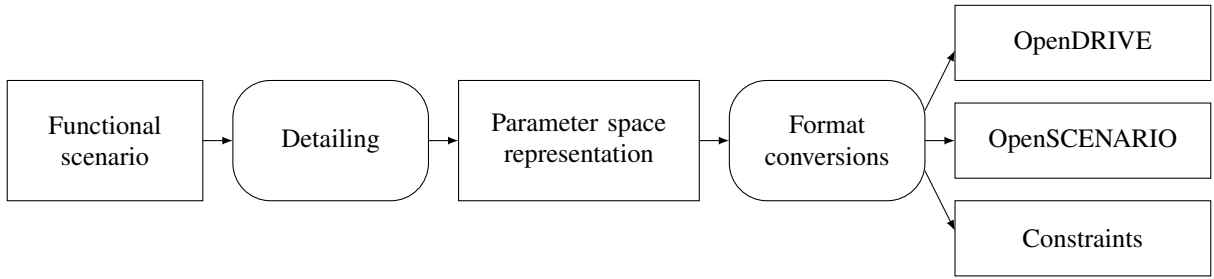

	\centering
	\include{process}
	\caption{Overview of the process for scenario transformation. Rectangles represent working products, rounded corners represent process steps.}
	\label{fig:process}
\end{figure*}

There are some publications which describe an event-based representation of scenarios, for example \cite{bach_model_2016} and \cite{xiong2013}.
Bach et al. \cite{bach_model_2016} describe a concept and a prototypic implementation for an abstract representation of concrete scenarios and, particularly, the interactions of traffic participants.
The concept is based on dividing each scenario into multiple acts and events.
Hence, the concept realizes a similar structure for the interactions of traffic participants as the data format OpenSCENARIO.
Xiong~\cite{xiong2013} presents an orchestration of scenarios in a simulation environment based on a knowledge base.
For this purpose, Xiong represents the sequence of actions and events in a scenario with the help of an ontology.
The ontology elaborately describes the elements of the road network and the interactions of the traffic participants as well as models (for example driver models) to be used.
The scenarios are executed in a simulation environment by a sequence control system and adjusted to the behavior of the driving function, if necessary.
Thus, this approach, which explicitly defines a vehicle under test, uses an ontology to control the other traffic participants to accomplish defined scenes in the scenario.
Hence, Bach et al. and Xiong present event-based notations for concrete scenarios respective test cases.
An approach for an automated detailing of a keyword-based scenario description to an event-based scenario description as a basis for a test case generation is neither described by Bach et al. nor by Xiong.
However, both publications can contribute to the structuring and the modeling of logical scenarios in the scope of virtual testing.

The concepts and implementations presented in this publication use functional scenarios generated by Bagschik~et~al.~\cite{bagschik_Szenariengenerierung_2018} as starting point for scenario detailing and transformation into the data formats OpenDRIVE and OpenSCENARIO.

%% file: process.tex
\usetikzlibrary{arrows}
\begin{tikzpicture}
\draw (-7.8,0) rectangle (-5.2,-1.6) node [pos=.5, align=center] {Functional\\ scenario};
\draw [rounded corners = 0.5cm] (-4.8,0) rectangle (-2.2,-1.6) node [pos=.5, align=center] {Detailing};
\draw (-1.8,0) rectangle (1.4,-1.6) node [pos=.5, align=center] {Parameter space\\ representation};
\draw [rounded corners = 0.5cm] (1.8,0) rectangle (4.4,-1.6) node [pos=.5, align=center] {Format\\ conversions};
\draw (4.8,1) rectangle (8.2,0) node [pos=.5, align=center] {OpenDRIVE};
\draw (4.8,-0.3) rectangle (8.2,-1.3) node [pos=.5, align=center] {OpenSCENARIO};
\draw (4.8,-1.6) rectangle (8.2,-2.6) node [pos=.5, align=center] {Constraints};

\draw [-latex](-5.2,-0.8) -- (-4.8,-0.8);
\draw [-latex](-2.2,-0.8) -- (-1.8,-0.8);
\draw [-latex](1.4,-0.8) -- (1.8,-0.8);

\draw [-latex](4.4,-0.4) -- (4.8,0.4);
\draw [-latex](4.4,-0.8) -- (4.8,-0.8);
\draw [-latex](4.4,-1.2) -- (4.8,-2);
\end{tikzpicture}

%% file: specification.tex
\section{Detailing a Keyword Based Scenario Description for Execution in a Simulation Environment}
\label{sec:specification}
As shown in Fig.~\ref{fig:process}, the transformation from a keyword-based scenario description into the formats for execution in a simulation environment is realized in two steps.
As starting point for the process, functional scenarios are used.
Each functional scenario is a semantic, keyword-based description and is modeled using the Web Ontology Language.
The generation of functional scenarios is described by Bagschik et al. \cite{bagschik_Szenariengenerierung_2018}.
 
Within the scope of detailing, the functional scenarios are imported and transformed into a parameter space representation.
For this, each term (like ``car'' or ``follow'') of the respective functional scenario is specified in detail through parameters. 
The  parameters are deduced by checking which parameters have to be defined in the data formats for simulation to represent the corresponding element.
Additionally, dependencies of the parameters as well as constraints for choosing parameter values are modeled.
These dependencies and constraints arise from physical relations of the parameters and interactions specified in the functional scenario as well as the requirements by the data formats for simulation. 

Within the scope of format conversion, for each data format the required information are extracted from the parameter space representation and converted into the syntax specified by the corresponding data format.
Furthermore, constraints for choosing parameter values are documented as set of rules in an additional file to provide following process steps (for example a test case generation) a basis for generating valid combinations of parameter values.
Afterwards, for each parameter, a default value is chosen according to the defined set of rules.
Thus, the converted scenarios specify the parameters to be varied, formalize their meaning for the execution in a simulation environment and define a valid default value for each parameter compliant with a documented set of rules.
Hence, the extracted scenarios build the basis for logical scenarios according to the definition of Menzel et al. \cite{menzel_process_2018}.
The generation of representative samples of concrete scenarios for each logical scenario according to a defined test concept is not part of this paper. 

\subsection{Scenario detailing}
As a basis for scenario detailing, a semantic scenario representation is used, which is modeled with the OWL data format.
This functional scenarios utilize a defined vocabulary as well as semantic relations between the terms of the vocabulary to describe the road level, the traffic infrastructure, the movable objects and the environmental conditions.
The vocabulary for the road level and the traffic infrastructure is defined by Bagschik et al.~\cite{bagschik_Szenariengenerierung_2018} in reference to the German standard how to construct freeways \cite{RAA}.
For the description of the interactions of the traffic participants, Bagschik et al.~\cite{bagschik_Szenariengenerierung_2018} use the maneuvers defined by Reschka~\cite{Reschka2016} based on T\"olle \cite{TolleFahrmanoverkonzeptfurmaschinellen1996} and Nagel and Enkelmann \cite{NagelGenericroadtraffic1991}.

\begin{figure}[]
	\centering
	\includegraphics[width=0.45\textwidth]{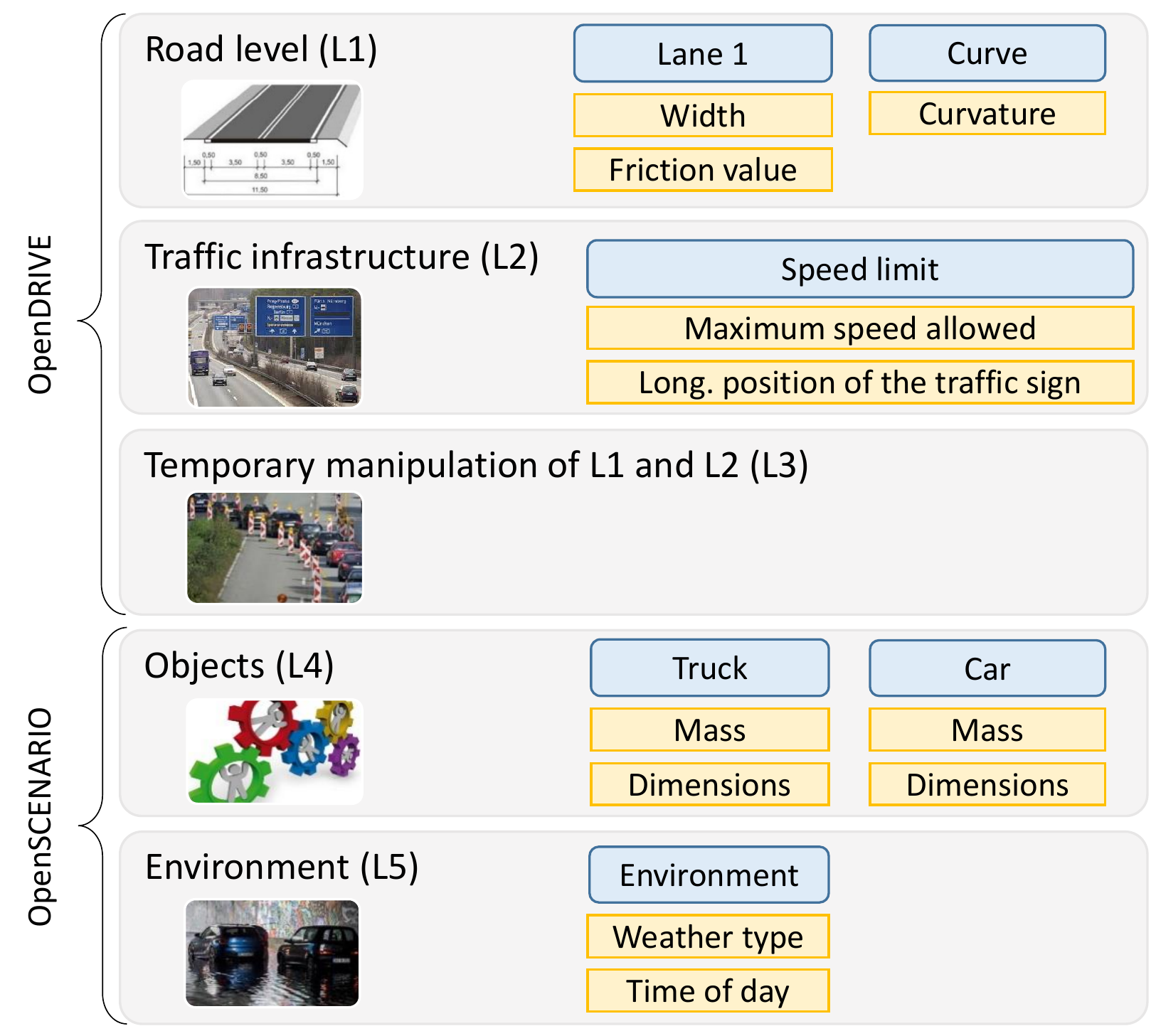}
	\caption{5-layer-model to structure scenarios by Bagschik~et~al.~\cite{bagschik_ontology_2018} based on Schuldt \cite{schuldt_beitrag_2017}. Blue boxes symbolize keywords from functional scenarios. Yellow boxes symbolize parameters according to the data formats for simulation.} 
	\label{fig:5-layer}
\end{figure}

In the first step of detailing, as shown in Fig.~\ref{fig:5-layer}, the terms used for scenario description are structured according to the 5-layer-model \cite{bagschik_ontology_2018}.
Afterwards, each term is augmented with parameters which specify the respective element in detail.
The required parameters for the description of each element depend on the data formats for the simulation environment.
Thus, within the scope of this paper, the parameters have been derived from the specifications of the data formats OpenDRIVE and OpenSCENARIO, which are used by the simulation environment Virtual Test Drive.
The data format OpenDRIVE describes the road network and, thus, the first three layers of the 5-layer-model.
The data format OpenSCENARIO describes the interactions of traffic participants as well as the environmental conditions and, thus, layers four and five of the 5-layer-model.

\begin{figure}[h]
	\centering
	\includegraphics[width=0.45\textwidth]{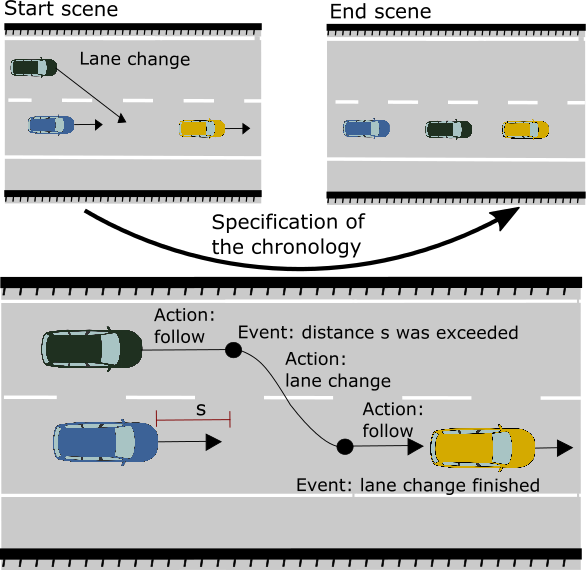}
	\caption{Specification of the chronology of interactions between traffic participants through an event-based representation.}
	\label{fig:events}
\end{figure}

In OpenSCENARIO, the trajectories of the traffic participants are described through actions (maneuvers and further commands for simulation control) and events, which trigger these actions. 
In functional scenarios, the movement of each vehicle is described through its relative position to other traffic participants in the start and in the end scene as well as the maneuver to be executed.
In order to represent the maneuvers defined in the functional scenarios with the help of parameters from OpenSCENARIO, the interactions of the traffic participants have to be transformed into an event-based representation, as shown in Fig.~\ref{fig:events}.
Hence, the maneuvers defined in the functional scenarios are specified in detail through additional design elements like actions and events.

In the second step of detailing, the relations and constraints between the elements and parameters of the elements are modeled.
Three categories of relations have been identified and are used for scenario description: arrangement relations, object dependencies and parameter dependencies.
These three types of relations are independent of each other and focus on different aspects of the scenario.
\begin{figure}[t]
	\centering
	\includegraphics[width=0.45\textwidth]{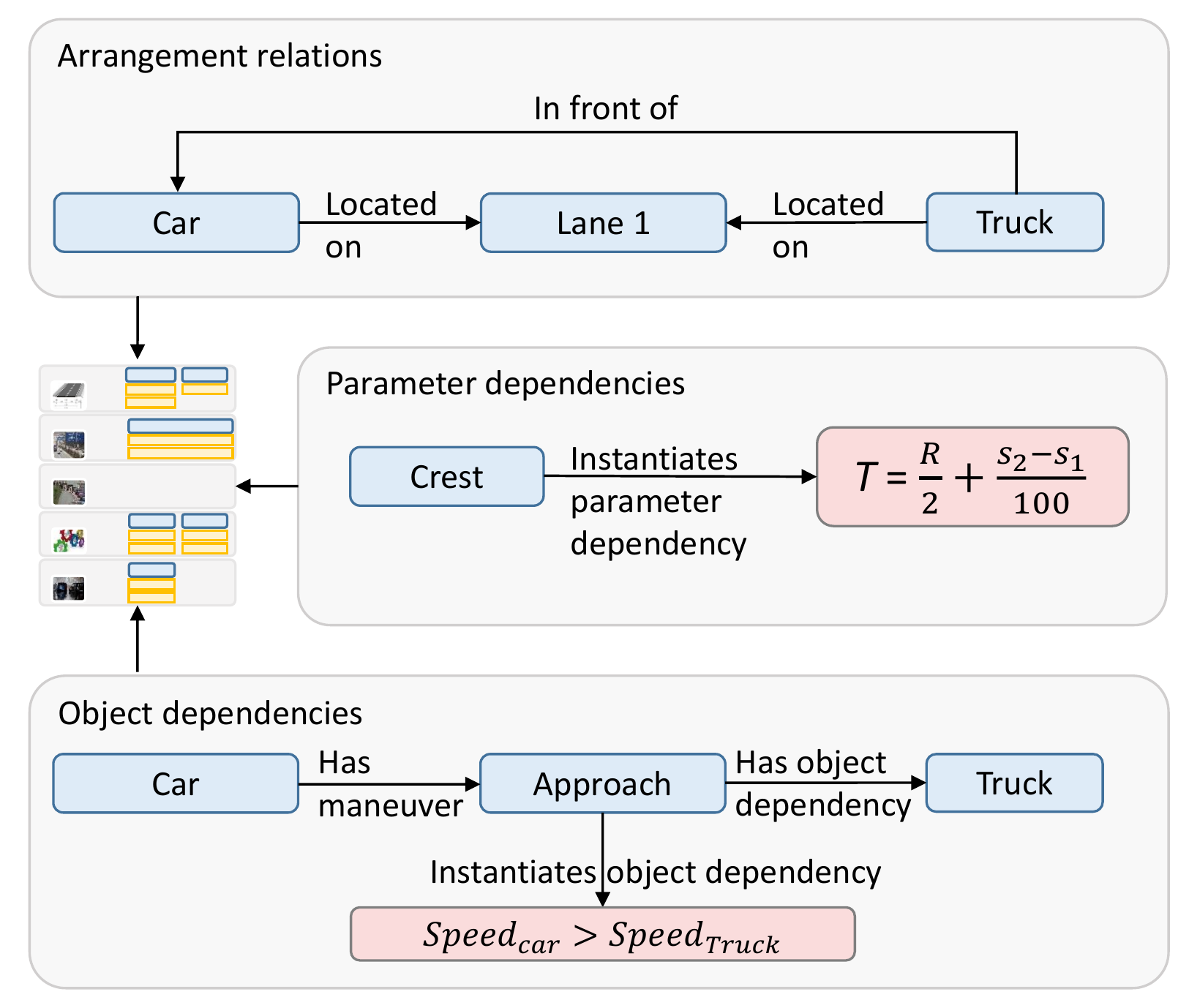}
	\caption{Augmentation of the parameter space with arrangement relations as well as object and parameter dependencies. The tangent length $T$ of a crest has to be calculated based on the radius $R$ as well as the initial tilt $s_1$ and the final tilt $s_2$.}
	\label{fig:relations}
\end{figure}

Arrangement relations characterize the relative positions of the elements to each other.
For this, all objects of the scenario are represented as nodes in a graph structure.
Adjacent or interacting objects are set into relation with the help of edges of the graph.
Among others, arrangement relations contain the mapping from lanes to road sections, the arrangement of lanes within a road section, the positions of the traffic participants on the road as well as the relative positions of the traffic participants to each other.
The vocabulary of functional scenarios by Bagschik et al. \cite{bagschik_Szenariengenerierung_2018} already describes arrangement relations so that this relations can be directly transformed.

Object dependencies describe dependencies between the objects of each scenario.
Each object dependency contains a set of rules, which describes constraints to be satisfied between the parameters of the linked objects.
These constraints are required to satisfy conditions by the functional scenarios or meet requirements by the data formats for the simulation environment.
For example, in functional scenarios described by Bagschik et al. \cite{bagschik_Szenariengenerierung_2018}, the interactions of traffic participants are described on a semantic level with the help of maneuvers.
To implement the maneuver approach on a parameter level, the speed of an approaching vehicle has to be faster than the speed of the vehicle ahead.
Additionally, in OpenDRIVE multiple lanes are structured as road section and put together to form a road network.
In consequence of this structure, successive lanes have to have the same lane width at their connection point.

Parameter dependencies describe dependencies between the parameters of one object.
Thereby, one parameter can be influenced by one or more parameters of the same object.
Parameter dependencies describe the conditions to be satisfied when choosing and combining parameter values.
These conditions result from standards or physical relations.
For example, in functional scenarios described by Bagschik et al. \cite{bagschik_Szenariengenerierung_2018} the elevation profile of a road is modeled using design patterns from the German standard how to construct freeways \cite{RAA} like planes, sag curve or crest curve.
The length of a crest curve cannot be chosen independently from the initial and final tilt, but has to be calculated according to the construction standard, as shown in Fig.~\ref{fig:relations}.
Hence, parameter dependencies help to choose parameter values of individual objects according to regulatory guidelines and physical dependencies.

\subsection{Format conversion}
Within the scope of format conversion, the information modeled in the parameter space representation are transformed into the data formats OpenDRIVE and OpenSCENARIO.
In the first step, the road representation with OpenDRIVE is generated.
For this, the elements of layers one to three of the 5-layer-model are transformed into the syntax of the OpenDRIVE format.
At the same time, a valid default value, which is calculated according to the constraints modeled in the parameter space representation, is assigned to each parameter.

In the second step, the representation of the traffic participants and environmental conditions in OpenSCENARIO is generated.
For this purpose, the elements of layers four and five of the 5-layer-model are transformed into the syntax of the OpenSCENARIO format and initialized with valid parameter values.
Thereby, the previously generated OpenDRIVE file is given as reference for the road representation.
Furthermore, the set of rules which was used for calculating default values for each parameter is documented in an additional file.
This way, the set of rules can be used in following process steps to generate valid combinations of parameter values, for example as part of a test case generation.

%% file: evaluation.tex
\section{Evaluation}
\label{sec:evaluation}
In this section, the generated OpenDRIVE and OpenSCENARIO files are examined.
Due to missing metrics as well as a joint set of scenarios for appliance and test, this examination formally is no evaluation, but gives a hint whether the scenarios are transformed correctly. 
For this, functional scenarios are generated according to the knowledge-based approach presented by Bagschik~et~al.~\cite{bagschik_Szenariengenerierung_2018}.
Afterwards, the generated scenarios are automatically transformed into a parameter space representation and converted into the formats OpenDRIVE and OpenSCENARIO using the tool presented in this paper.
The test scenarios are selected and evaluated separately for both formats.

\subsection{OpenDRIVE}
To describe the road network, the vocabulary of functional scenarios includes multiple variations of elevation profiles (like incline or decline), alignments (like straight or curve) and lane types (like hard shoulder or driving lane).
Furthermore, various elements of the traffic infrastructure (like traffic signs or guard rails) are described.
Due to various combinations of these elements, a large number of different road sections can be generated.
In the current implementation of the knowledge-based scenario generation by Bagschik~et~al.~\cite{bagschik_Szenariengenerierung_2018}, the road network of each functional scenario only consists of one road section.
A combination of successive road sections to longer roads is currently not implemented and, thus, limits the variety of input data for evaluation.

\begin{figure} 
	\centering
	\subfloat[OpenDRIVE track displayed in OpenDRIVE-Viewer\label{1a}]{%
		\includegraphics[width=0.49\linewidth]{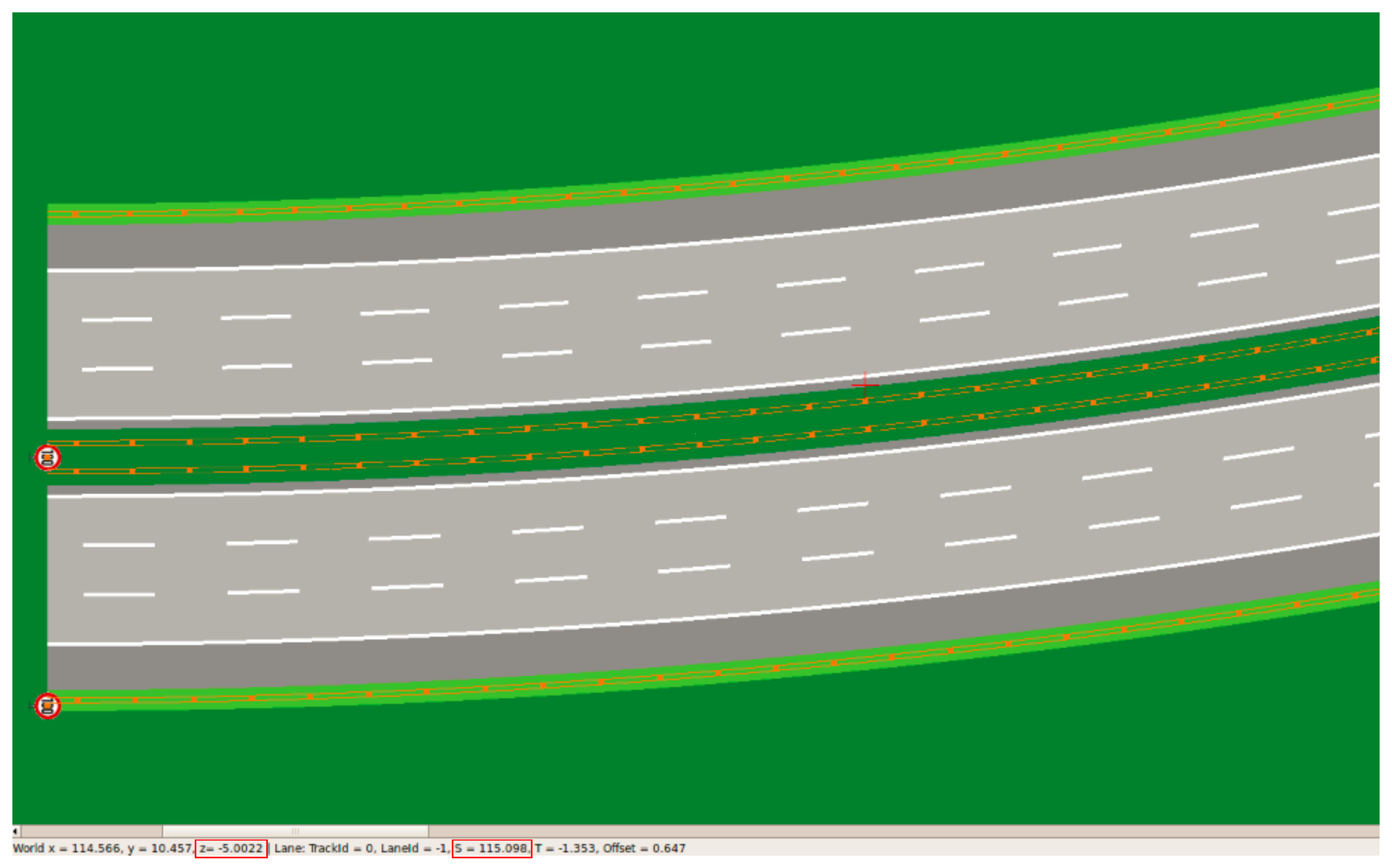}}
	\hfill
	\subfloat[OpenDRIVE track displayed in Virtual Test Drive\label{1b}]{%
		\includegraphics[width=0.49\linewidth]{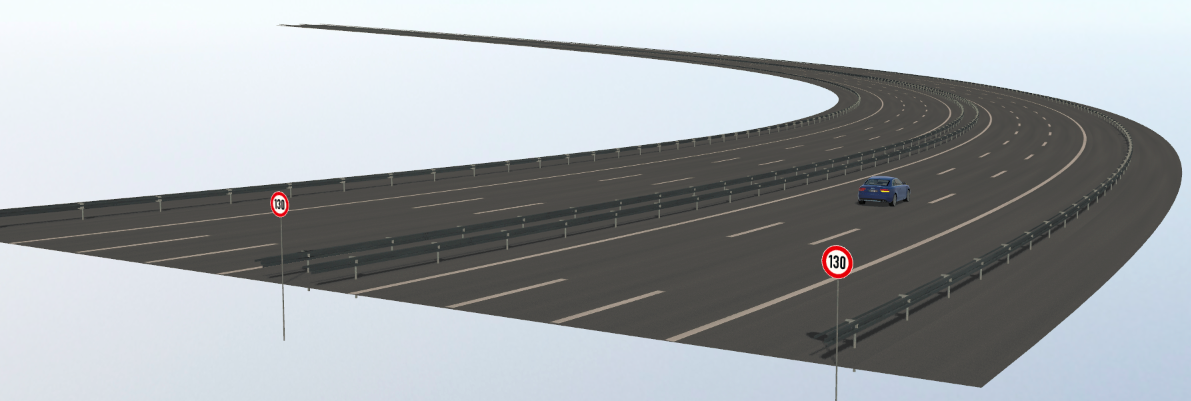}}
	\caption{Three lane freeway in a curve with decline, speed limit and guard rail displayed in OpenDRIVE-Viewer (\ref{1a}) and Virtual Test Drive (\ref{1b}).}
	\label{fig:eval} 
\end{figure}
Within the OpenDRIVE format, there only exist dependencies between successive road sections and lanes.
The design elements of a road section, like elevation profile and road geometry, can be specified independently.
As each road network of the generated functional scenarios only consists of one road section, not all possible combinations have to be tested.
For evaluation it only has to be tested whether each single vocable for the road description is mapped correctly (according to the classes and attributes required in the OpenDRIVE format) to the OpenDRIVE data format.
Hence, for evaluation all in all five scenarios have been chosen, which make use of the complete vocabulary for road description.

The OpenDRIVE files were evaluated in two steps.
In the first step, the generated roads have been examined for implementation according to the specified functional scenarios as well as for parametrization compliant with the standards.
For example, the geometry (alignment and elevation profile) was evaluated by checking the course of coordinates with the help of the OpenDRIVE-Viewer (see Fig. \ref{1a}).
In the second step, the road description was loaded in Virtual Test Drive (see Fig. \ref{1b}) and examined whether all 3D-graphics were displayed and all lanes were linked correctly.
Hence, a virtual vehicle was defined to drive along the modeled course.
With this procedure has been shown that the parameter space for road description is generated according to the respective functional scenario as well as the given standards and is converted correctly into the OpenDRIVE data format.

\subsection{OpenSCENARIO}
The vocabulary of functional scenarios for describing traffic participants includes various vehicle types (like car or truck) and maneuvers (like lane change or approach).
The positions of the vehicles are linked with each other with the help of a lane relative position grid.
Furthermore, various environment features are described with the help of different weather conditions (like rain or clear sky) and different times of day (like morning or midday).

The OpenSCENARIO files were evaluated in two steps as well.
In the first step, the generated OpenSCENARIO files were checked against a schema file in a static test.
In this scope, the syntax of each OpenSCENARIO file was compared to the formal specification of the OpenSCENARIO data format. 
This way, missing elements and attributes can be identified.
For the static test with the schema file, six scenarios have been chosen, which include all OpenSCENARIO design elements that are needed to implement the vocabulary of functional scenarios generated by Bagschik~et~al.~\cite{bagschik_Szenariengenerierung_2018}.
The checks against the schema file showed that the functional scenarios are converted into OpenSCENARIO files according to the data format specification.
However, such static tests do not reveal whether all elements defined in a functional scenario are implemented in the respective OpenSCENARIO file and initialized with valid default values.

Thus, in the second step, the OpenSCENARIO files are loaded and executed in a simulation environment.
Within this scope, it is checked whether each generated OpenSCENARIO file is executed in simulation according to the respective functional scenario.
During this test, the main questions are:
\begin{itemize}
	\item Do all vehicles start at the position defined in the start scene?
	\item Does the behavior of each vehicle comply to the maneuver defined in the start scene?
	\item Do all maneuvers start and end at the correct point in time so that no crashes are caused? 
	\item Do all vehicles end at the positions defined in the end scene after they have executed their maneuvers?
\end{itemize}

To create input data for the test, the authors generated more than 10,000 functional scenarios utilizing the approach described by Bagschik~et~al.~\cite{bagschik_Szenariengenerierung_2018}.
This vast number of scenarios cannot be evaluated manually.
Hence, the authors defined nine test goals and picked relevant scenarios to be tested based on these nine goals.
Every test goal describes one part of the program for detailing and converting scenarios (like the calculation of start speeds for the traffic participants), which should be checked explicitly.
Additionally, each test goal defines one test sketch, which describes the rough structure and storyboard of scenarios which can be utilized to check the respective test goal.
In total, 30 scenarios have been selected according to the test goals and executed in the simulation environment correctly (as specified in the respective functional scenario).

%% file: conclusion.tex
\section{Conclusion and outlook}
\label{sec:conclusion}
In this paper, the authors presented an approach to transform a keyword-based scenario description into the formats for simulation.
For this purpose, each keyword-based scenario is transformed into a parameter space representation and afterwards converted into the data formats OpenDRIVE and OpenSCENARIO.
The automatically generated scenarios have been evaluated and executed in simulation according to the specified functional scenarios.

The knowledge-based scenario generation by Bagschik et al. \cite{bagschik_Szenariengenerierung_2018} and the presented approach enable a semantic variation of scenarios and build the basis for logical scenarios.
In combination with a data-driven approach, these logical scenarios can be used as a basis for clustering scenarios from measurement data.
Furthermore, these logical scenarios could build a reference for scenario variety and, thus, ensure a wide semantic variation of scenarios.

Nevertheless, there are some limitations to the presented approach.
First of all, both in the functional scenarios and the generated data files for simulation only vehicles which are controlled via simulation are defined.
The specification of a vehicle which is controlled via an automated driving function has to be added in the future.
The degree of freedom resulting from the undefined behavior of the system under test could lead to an unwanted sequence of maneuvers and interactions.
These possible action alternatives have to be considered in the sequence control for simulation and, thus, have to be described in the data files for simulation.

Additionally, in the current implementation, the knowledge about detailing keywords to a parameter space and deducing constraints from keywords is modeled directly in the source code.
Within the scope of a traceable scenario generation along the development process, this knowledge should be modeled explicitly, for example as an ontology.
This gets even more important, when the scenarios get more complex (rising number of relations and constraints).
As stated above, this would be the case, when defining a system under test or specifying scenarios in a more complex domain, like urban areas.

Another possible limitation is the definition of constraints, which ensure that the generated scenarios are physically reasonable.
Those constraints have to be carefully selected to guarantee that only irrelevant scenarios are excluded.

\addtolength{\textheight}{-11.5cm}
\newpage